# POLARIZATION MEASUREMENTS - A NUMERICAL APPROACH

Aleksandar Gjurchinovski [*]

*Department of Physics, Faculty of Natural Sciences and Mathematics, Sts. Cyril and Methodius University, P.O.Box 162, 1000 Skopje, Macedonia*

**ABSTRACT**

We developed a specific numerical fitting procedure, based on the least squares method, to calculate the parameters of the polarization ellipse by only using a single polarizer and a detection device for measuring the outgoing intensities.

**PACS Numbers:** 42.25.Ja, 02.60.Ed, 02.10.Ud

## I. INTRODUCTION

Polarization is a physical property common to all types of vector waves. It refers to the time behavior of one of the field vectors describing that particular wave, observed at some fixed point in space. In this case, we will be concerned with the vibration of the electric field vector **E**. Polarization measurements can be quite useful in a variety of situations [1]. For example, an interesting case occurs when a plane polarized light beam (a light-wave probe) is allowed to pass through a succession of optical devices, each of which produces a specific change only in the state of the polarization of the probe. Knowledge of the initial and final states of polarization of the probe can be used to investigate the assemblage that modifies the state of polarization. The description of the optical system that interacts with the light-probe, as well as the effect of modifying the polarization state, can be done with the Jones matrix calculus. However, in order to preserve an intuitive understanding of the physical processes involved in the polarization measurements, we will not use the Jones calculus here. We emphasize that the equation for the transmitted intensity derived further in the text can also be achieved with the Jones matrix method. For more detailed discussion, the reader is referred to the literature [1,2,3,4].

Our task will be to develop a practical procedure for determining the parameters of the polarization ellipse of an arbitrary beam of plane polarized monochromatic light. In doing so, we are only allowed to use a polarizer and a detection device for measuring the intensity of the light transmitted through the polarizer. According to the Stokes parameters [1,5], measurements performed for three incompatible orientations of the transmission axis of the polarizer are enough to find the shape and the orientation of the polarization ellipse (for an ideal polarizer, $\psi_1$ and $\psi_2$ are compatible angles if $\psi_1 = \psi_2 \pm 180º$, where by $\psi$ we designate the angle that describes the orientation of the transmission axis of the polarizer). But practice shows that the results are more accurate and reliable when doing measurements for a large number of angles.

---

[*] Electronic address: agjurcin@iunona.pmf.ukim.edu.mk

## II. THE LEAST SQUARES APPROXIMATION

When performing polarization measurements on a plane polarized light by using only one polarizer and a detection device for measuring the outgoing intensity (Fig. 1), it is appropriate to calculate several points for determining the polarization ellipse. This is due to the fact that small fluctuations of the intensity and imperfections of the polarizers can lead to huge deviations in the values of the parameters of the ellipse. Our mathematical derivations can be further simplified by taking only one assumption - the **E** vector is rotating in a plane and it describes a closed curve - an ellipse.

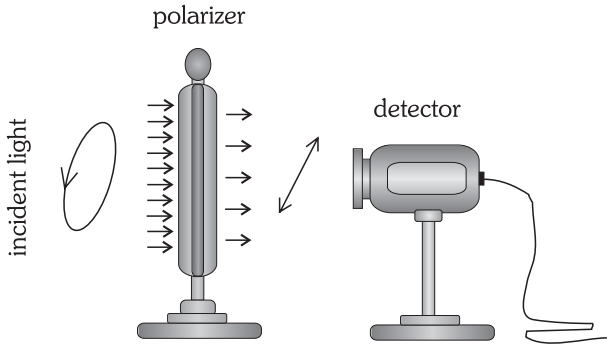
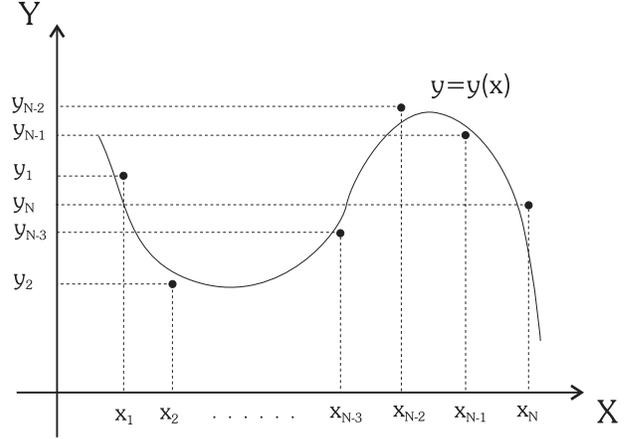

**Fig. 1.** Setup for measuring the intensity of the light transmitted through the polarizer.

**Fig. 2.** The least squares approximation. The curve $y = y(x)$ is the best fit of the experimental data

The algorithm for fitting the experimentally obtained intensities for several values of the angle of the polarizer can be accomplished with the least squares approximation [6]. The mathematical structure of this method can be explained by using the graph depicted in Fig. 2. We designate by $x_1, x_2, x_3, ..., x_N$ the measured values of the physical observable $x$, and by $y_1, y_2, y_3, ..., y_N$ the measured values of the physical observable $y$ corresponding to $x$. In our case, $x$ is the angle $\psi$ that describes the orientation of the transmission axis of the polarizer, and $y$ is the intensity $I$ of the outgoing light. Here N denotes the total number of performed measurements, and $N \geq 3$. The aim of the least squares method is to fit the experimental points $(x_i, y_i)$ to a theoretical curve $y = y(x, a_1, a_2, ..., a_K)$, or in other words, to adjust a set of parameters $\{a_j ; j = 1, 2, ..., K\}$ in a way which will cause the function $y = y(x, a_1, a_2, ..., a_K)$ to become "the closest one" to the measured points $(x_i, y_i)$ in the sense that the value of

$$\Delta = \sum_{i=1}^{N} \left[ y(x_i, a_1, a_2, ..., a_K) - y_i \right]^2 \qquad (1)$$

is minimized. Usually, $y = y(x, a_1, a_2, ..., a_K)$ is called *trial function* (or *trial curve*), and it corresponds to the equation obtained with the theoretical model which gives a proper description of the physical situation. The least squares approximation is then given through the following set of K equations in K unknowns $\{a_j\}$:

$$\frac{\partial}{\partial a_j} \Delta(a_1, a_2, ..., a_K) = 0 \quad , \quad j = 1, 2, ..., K. \qquad (2)$$

The system (2) is usually nonlinear, and, if the solution of the system exists and if it is unique, it can be approximately solved by using certain numerical techniques. When the set of values $\{a_j\}$ are found, the least squares curve can be written as $y = y(x)$.

## III. THE POLARIZATION ELLIPSE

Before we proceed, we need to make some remarks concerning the **E** - field of the elliptically polarized light which is incident on the polarizer. Strictly speaking, the endpoint of the **E** vector is actually describing a periodic helix-like curve along the direction of the Poynting vector in three dimensional space. The period of the helix equals that of the light wave (for a visible light, $T \sim 10^{-15}$ s), and the projection of the helix on the plane perpendicular to the Poynting vector is an ellipse. The shape and the orientation of the ellipse characterize the polarization state of the light, which, on another hand, can be conveniently described by introducing certain parameters [1,4,5]. In Fig. 3 we have a *XY* drawing of an **E** - ellipse. It must be emphasized that the light beam is perpendicular to the plane of the figure and its direction of propagation is towards the reader. The same holds for Fig. 4.

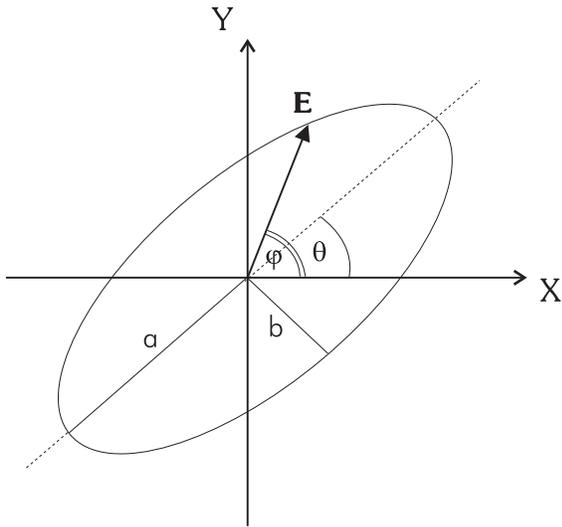
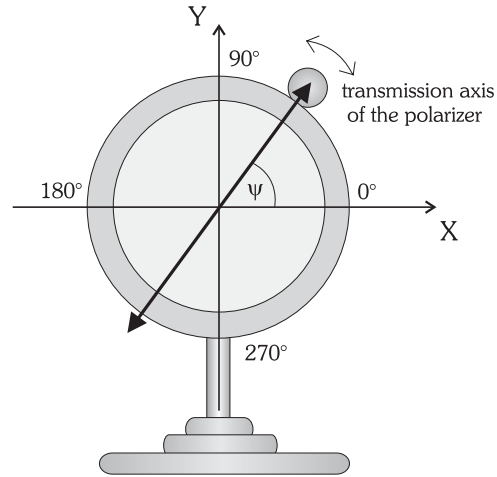

**Fig. 3.** Polarization ellipse of the plane polarized light which is incident on the polarizer. The electric field vector **E** changes its direction and its magnitude while rotating in plane.

**Fig. 4.** Back view of the polarizer. The transmitted light beam is perpendicular to the plane of the figure and it propagates towards the reader, while its electric field vector $\mathbf{E}_\psi$ is oscillating along the direction of the transmission axis.

The *inclination angle* $\theta$ (or *azimuthal angle*, *azimuth*) is the angle between the major axis of the ellipse and the positive direction of the *X* - axis, and defines the orientation of the ellipse in its plane [1]. All physically distinguishable azimuths can be obtained by limiting $\theta$ to the range $-\frac{\pi}{2} \leq \theta < \frac{\pi}{2}$. The *ellipticity* $\varepsilon = \frac{b}{a}$ is the ratio of the length of the semi-minor axis of the ellipse $b$ to the length of its semi-major axis $a$. The *amplitude W*, also known as the *size* of an elliptical vibration, is defined as

$$W = \sqrt{a^2 + b^2} \:. \tag{3}$$

The amplitude *W* is a measure of the strength of the elliptical vibration and its square is proportional to the energy density of the wave at the point of observation of the field. The *phase difference* δ between *X* - and *Y* - oscillation of the electric field vector **E** satisfies the following relation [4]

$$\cos\delta = \frac{E_{X\max}^2 - E_{Y\max}^2}{2E_{X\max}E_{Y\max}}\tan 2\theta. \qquad (4)$$

It will be shown later in the text that $E_{Xmax}$ and $E_{Ymax}$ are correlated with *a*, *b* and θ. The *handedness* of the polarization ellipse determines the way in which the ellipse is described. It is a parameter that can assume only one of two discrete "values". The polarization is *right handed* if the ellipse is traversed in a clockwise sense when looking against the direction of propagation of the light beam. The polarization is *left handed* if the ellipse is traversed in a counterclockwise sense. It follows that the actual shape, size and handedness of the **E** - ellipse determine the state of polarization of the incoming light.

The equation of the **E** - ellipse in polar coordinate system (*E*,φ) can be expressed as

$$E(\varphi) = \sqrt{\frac{a^2 b^2}{a^2 \sin^2(\varphi-\theta) + b^2 \cos^2(\varphi-\theta)}}. \qquad (5)$$

In the process of derivation of the fitting algorithm we assume that the time dependence of the angular displacement φ = φ(*t*) of the **E** - field is not known.

The electric field vector **E**$_\psi$ of the linearly polarized light which is transmitted by the polarizer is an orthogonal projection of the **E** - field of the incoming elliptically polarized light on the direction of the transmission axis ψ of the polarizer

$$\mathbf{E}_\psi(t) = E[\varphi(t)] \cdot \cos[\varphi(t) - \psi] \cdot \mathbf{e}_\psi, \qquad (6)$$

where **e**$_\psi$ is the unit vector in the ψ direction. Taking into account that $I = |\mathbf{E}|^2$, we obtain the intensity of the transmitted light

$$I_\psi(t) = \frac{a^2 b^2}{a^2 \sin^2[\varphi(t)-\theta] + b^2 \cos^2[\varphi(t)-\theta]} \cdot \cos^2[\varphi(t)-\psi]. \qquad (7)$$

It can be noticed that $I_\psi(t)$ is time-dependent and it oscillates between two extreme values. The maximal value of the intensity which passes through the polarizer when the orientation of its transmission axis is ψ degrees counterclockwise towards the *X* - axis (Fig. 4), is described by

$$I(\psi) = a^2 \cos^2(\psi-\theta) + b^2 \sin^2(\psi-\theta). \qquad (8)$$

In order to derive equation (8), one must take the first derivative of (7) with respect to φ and then equal the result to zero. By solving the equation obtained in this manner, one gets the optimal value of φ. It can be shown that there are two kinds of solutions in φ. When the first solution is replaced in (7), one gets the minimal value of the transmitted intensity. The minimal value of the intensity is zero, which is quite obvious. When the second solution in φ is replaced, one obtains the relation (8).

Now, the usual measuring equipment, which can be found in almost every optical laboratory, consists of a photodiode connected to a multimeter via amplification device. The intensity value of the light which is detected by the multimeter is actually the *root-mean-square* (*RMS*, or, *effective value*) of the transmitted intensity averaged for one period (the time it takes for the **E** vector to describe the ellipse once, rotating in the polar plane for the angle of $2\pi$ radians, or, equivalently, the time required for the $\mathbf{E}_\psi$ vector to make one complete linear oscillation along the direction defined with $\psi$)

$$I_{RMS}(\psi) = \sqrt{\frac{1}{T}\int_0^T I_\psi^2(t)dt}\ . \tag{9}$$

However, if we connect the photodiode to an oscilloscope or some other device from which we can read the maximal value of the transmitted light intensity, we can apply equation (8) as a trial curve in the fitting procedure. Otherwise, there should be a real positive function in the equation (8), denoted by $\alpha(\psi)$, which multiplies the whole right hand side of it. The value of $\alpha(\psi)$ can be theoretically derived from (9) and from (7) if the time dependence of the angular displacement $\varphi = \varphi(t)$ is known. Immediately, it follows that $\alpha(\psi) = \dfrac{I_{RMS}(\psi)}{I(\psi)}$. In the best case, $\alpha(\psi)$ will not be dependent on the orientation of the transmission axis. In that case, $\alpha$ is a real positive constant, called *scaling* (or *amplification*) *factor*. This allows us to use equation (8) in its original form while thinking of the whole process of measuring the intensity like it is performed with an imaginary amplifying device with an amplification factor of $1/\alpha$. The only value that is affected by this is the amplitude $W$, but the shape and the handedness of the ellipse remain unchanged.

While further developing the algorithm, we assume that the measurements are performed with an apparatus which detects the maximal intensity value. Then, equation (8) can be implemented as a trial curve without any modifications.

## IV. FITTING PROCEDURE

The next step is obtaining an algorithm for fitting the $I(\psi)$ curve. The reader can notice that in order to calculate the parameters of the polarization ellipse described by the $\mathbf{E}$ - field, we are actually fitting the intensity curve of the transmitted light. The trial curve we will use in the least squares fitting procedure is

$$I(\psi) = a_1^2 \cos^2(\psi - a_3) + a_2^2 \sin^2(\psi - a_3)\ . \tag{10}$$

The meaning of the parameters $a_1$, $a_2$ and $a_3$ is obvious, if one compares (10) with (8):

$$a_1 = a\ ,\ a_2 = b\ ,\ a_3 = \theta.$$

In the following, we will implement the least squares method to adjust the parameters $a_1$, $a_2$ and $a_3$. The $\Delta$ function, which is expressed by (1), in our case obtains the following form

$$\Delta = \sum_{i=1}^N \left[a_1^2 \cos^2(\psi_i - a_3) + a_2^2 \sin^2(\psi_i - a_3) - I_i\right]^2\ . \tag{11}$$

By $I_i$ we denote the $i$ - th maximal intensity value measured when the angle of the transmission axis of the polarizer is $\psi_i$ degrees counterclockwise towards the positive direction of the $X$ - axis. Applying equations (2) for $a_1$, $a_2$ and $a_3$, and after some algebraic manipulations, one comes to the following set (or system) of linear equations in $p$, $q$ and $r$ :

$$\begin{cases} D_{11}\ p + D_{12}\ q + D_{13}\ r & =\ F_1 \\ D_{21}\ p + D_{22}\ q + D_{23}\ r & =\ F_2 \\ D_{31}\ p + D_{32}\ q + D_{33}\ r & =\ F_3 \end{cases},\tag{12}$$

where with *p*, *q* and *r* we substituted the expressions

$$p = \frac{a_1^2 + a_2^2}{2}, \quad q = \frac{a_1^2 - a_2^2}{2}\cos(2a_3), \quad r = \frac{a_1^2 - a_2^2}{2}\sin(2a_3). \quad (13)$$

The values of the coefficients $D_{nm}$ and $F_n$ of the linear system (12) are:

$$D_{11} = N, \; D_{12} = D_{31} = \sum_{i=1}^{N}\cos(2\psi_i), \; D_{23} = \sum_{i=1}^{N}\sin^2(2\psi_i), \; D_{32} = \sum_{i=1}^{N}\cos^2(2\psi_i), \; D_{13} = D_{21} = \sum_{i=1}^{N}\sin(2\psi_i),$$

$$D_{22} = D_{33} = \sum_{i=1}^{N}\sin(2\psi_i)\cos(2\psi_i), \; F_1 = \sum_{i=1}^{N}I_i, \; F_2 = \sum_{i=1}^{N}I_i\sin(2\psi_i), \text{ and } F_3 = \sum_{i=1}^{N}I_i\cos(2\psi_i).$$

It can be noticed that by implementing (13), one is assured that the system (12) becomes linear.

Although derivation of the least squares equations for the ellipticity ε, the amplitude *W* and the azimuth θ is straightforward from (13) and the definitions given at the beginning of the previous section (we take into account that $a = a_1$, $b = a_2$ and $\theta = a_3$), deriving the equation for the phase difference requires some explanation. Bearing in mind that the maximal intensity equals the square modulus of the amplitude, the values of $E_{Xmax}$ and $E_{Ymax}$ follow from equation (8) by putting ψ = 0 for $E_{Xmax}$ and ψ = π/2 for $E_{Ymax}$,

$$E_{X\max} = \sqrt{a^2\cos^2\theta + b^2\sin^2\theta}, \quad (14.a)$$

$$E_{Y\max} = \sqrt{a^2\sin^2\theta + b^2\cos^2\theta}. \quad (14.b)$$

Then, substituting (14.*a*) and (14.*b*) in (4), and after some rearrangements, one obtains the least squares relation for the phase parameter. Finally, the equations for the parameters of the ellipse of polarization are found to be correlated to *p*, *q* and *r* in the following manner

Ellipticity $$\varepsilon = \sqrt{\frac{p - \sqrt{q^2 + r^2}}{p + \sqrt{q^2 + r^2}}}, \quad (15)$$

Amplitude $$W = \sqrt{2p}, \quad (16)$$

Azimuth $$\tan(2\theta) = \frac{r}{q}, \quad (17)$$

Phase $$\cos\delta = \frac{r}{\sqrt{p^2 - q^2}}. \quad (18)$$

## V. CONCLUDING REMARKS

The least squares fitting procedure for calculation of the polarization ellipse parameters is based on the equations (15) - (18) and the system (12). The algorithm can be carried out in several steps:

1. Perform a measurement on the maximal intensity of the light $I$ which passes through the polarizer. Repeat this for N different values for the orientation of the angle of the transmission axis $\psi$. Take into account that the values $\psi_i$ must not be compatible to each other and that you need to perform three or more measurements for determining the ellipse correctly.

2. When the N pairs of real numbers $(\psi_i, I_i)$ are established, calculate the coefficients $D_{nm}$ and $F_n$ ($n, m = 1, 2, 3$) of the linear system (12).

3. By using the standard procedures for solving linear systems (Row-reduction method or Cramer's rule [7]), find the values of $p$, $q$ and $r$.

4. By substitution of the values of $p$, $q$ and $r$ in the equations (15) - (18), get the parameters of the ellipse. Make a graph of the polarization ellipse. Discuss the result.

Due to the periodicity of the tangens function, there is a problem in determining the inclination angle $\theta$. The reader should recall that since $\tan(2\theta) = \tan(2\theta + \pi)$, there are two azimuths for which the equation (17) is satisfied: $\theta_1 = \frac{1}{2}\arctan\left(\frac{r}{q}\right)$ and $\theta_2 = \frac{1}{2}\arctan\left(\frac{r}{q}\right) + \frac{\pi}{2}$. Correct azimuth can be obtained by calculating the value of $\Delta$ function given with equation (11), once for $a_3 = \theta_1$, and once again for $a_3 = \theta_2$. In both cases, $a_1$ and $a_2$ can be expressed through $p$, $q$ and $r$ by eliminating $a_3$ from the set of equations (13). The correct value of the azimuth is the one that makes the value of $\Delta$ minimal.

However, the handedness of the ellipse, or in other words, the sense in which the ellipse is described by the **E** vector (clockwise or counterclockwise), cannot be evaluated. This stems on the fact that cosine function is even, so the sign of $\delta$ cannot be determined from (18). For its evaluation, further experiments must be performed, for example, by adding a quarter-wave retardation plate to the setup [4].

**ACKNOWLEDGMENTS**

This article is based on certain theoretical explorations of the optical measurements performed during the author's visiting of the Department of Electronics and Information Systems (ELIS) at the University of Gent, Belgium, in the framework of TEMPUS JEP N°. 13576-98. The author would like to thank the LCD Research Group, particularly Prof. D-r. Kristiaan Neyts, D-r. Herbert De Vleeschouwer, D-r. Stanislaw Rozanski, Stefaan Vermael and Chris Desimpel for useful suggestions concerning the article. The author would also like to acknowledge Prof. D-r. Viktor Urumov and Prof. D-r. Hendrik Ferdinande for successfully managing the whole project.